\documentstyle[12pt]{article}
\date{}
\begin{document}

\title{The Baryon Spectrum and Chiral Dynamics}

\author{L. Ya. Glozman$^{1,2}$ and D.O. Riska$^1$}
\maketitle

\centerline{\it $^1$Department of Physics, University of Helsinki,
00014 Finland}

\centerline{\it $^2$Alma-Ata Power Engineering Institute,
480013 Alma-Ata, Kazakhstan}

\setcounter{page} {0}
\vspace{1cm}

\centerline{\bf Abstract}
\vspace{0.5cm}

The fine structure of the
low energy part of the nucleon and strange hyperon spectra, which
are formed of single states without parity doublets, may be
understood in terms of an $SU(3)$ flavor-symmetric
quark-quark interaction that describes chiral pseudoscalar boson
exchange. The model predicts the fine structure splittings
within 10-30\% of their empirical values
and provides an
explanation of the reversed ordering of the lowest positive
and negative parity resonances in the nucleon and the $\Lambda$
hyperon spectra.\\

PACS numbers 12.39.-x, 12.39.Fe

Submitted to Physical Review Letters

\vspace{5cm}

\noindent
\small
Research Institute for Theoretical Physics, University of Helsinki,\\
Preprint HU-TFT-94-47, Nov. 11, 1994
LANL hep-ph 9411279

\newpage
\normalsize
The spectra of the nucleons and the strange hyperons separate into a
low energy sector formed of single states without parity partners
and a high energy sector formed
of near degenerate parity doublets, the latter feature being
particularly evident in the spectrum of the $\Lambda$ hyperon.
This is interpreted as a
consequence of the (approximate) chiral
symmetry of QCD being realized in the
hidden (Nambu-Goldstone) mode in the low energy
(and low temperature) sector, whereas it is realized in the
explicit (Wigner-Weyl) mode in the high
energy (and high temperature) sector. Instead of by parity doubling
of the spectrum the hidden mode of chiral
symmetry is revealed by the existence of
the nonet of low mass pseudoscalar (Goldstone)
bosons and constituent quarks.\\

This "chiral" pseudoscalar  octet (the $\eta'$ decouples because
of the $U(1)$ anomaly [1]) will mediate interactions between the
constituent quarks and thus leads to
fine structure in the baryon spectrum, the gross features of
which are caused by the confining interaction.
The simplest representation of the interaction
that is mediated by the chiral octet would be

$$H_\chi\sim -C_\chi\sum_{i<j}\vec \lambda^F_i \cdot \vec \lambda^F_j\,
\vec
\sigma_i \cdot \vec \sigma_j.\eqno(1)$$
Here the $\{\vec \lambda^F_i\}$:s are flavor $SU(3)$ matrices and the
$i,j$ sums run over the constituent quarks. The form of this
interaction is an immediate generalization of the spin-spin
component of the pseudoscalar (pion) exchange interaction, the tensor
component of which in the present context is insignificant. The
coefficient $C_\chi$ represents an averaged radial matrix
element.
A refined version of the interaction
(1) would contain a $SU(3)_F$ flavor symmetry breaking term
to account for the mass splitting within the pseudoscalar octet,
that arises from the explicit chiral symmetry breaking in QCD.\\

The chiral field interaction should be contrasted in form with the
color-magnetic interaction [2]

$$H_c\sim-\alpha_s\sum_{i<j} \vec \lambda_i^C\cdot
\vec \lambda_j^C \vec
\sigma_i\cdot \vec \sigma_j\delta(\vec r_{ij}),\eqno(2)$$
where the $\{\vec \lambda_i^C\}$:s are color $SU(3)$ matrices, and
which should
be important in the region of explicit chiral symmetry at short
distances and high energy. It is in fact this color-magnetic
interaction, which has been used in earlier attempts to describe the baryon
spectra with the constituent quark model [3,4]. Although many of the
qualitative and some of the quantitative features of the
fine structure of the baryon
spectra can be described by the interaction (2) a number
of outstanding features have proven hard to explain in this approach.
The most obvious one of these is the different ordering of the
(confirmed) positive and negative parity resonances in the
spectra of
the nucleon and the $\Sigma$ hyperon
on the one hand and the $\Lambda$
hyperon on the other, and in particular the difficulty in describing the
low mass of the
$\Lambda(1405)$ resonance. A second such feature
is the absence of empirical
indications
for the spin-orbit interaction
that should accompany the
color-magnetic interaction (2). We shall show below that the chiral
pseudoscalar interaction (1) provides a simpler description of the
fine structure of the low
energy baryon spectra, that automatically implies the reversal of the
ordering of the even and odd parity states between the nucleon and
$\Lambda$ hyperon spectra. Moreover we show that when the
effective coupling
strength $C_\chi$ in (1) is determined by the
the average splitting between baryon states that have the same
orbital and spin-flavor symmetry
a quite satisfactory description of the fine structure
of the whole low lying baryon spectrum is
achieved already in lowest order. Finally no spin-orbit force problem
appears as pseudoscalar interactions have no spin orbit component.
This then suggests that it is the chiral field interaction (1),
which plays the dominant role in ordering the baryon spectrum in
the region of hidden chiral symmetry.\\

Consider first the
lowest 3 $J=\frac{1}{2}$ baryon states with strangeness 0 and -1,
which are listed in Table 1.
The symmetry classification of these states is denoted by
the Young pattern
$[f]$, which is the permutational symmetry with respect to
the relevant transformation group. The color singlet structure
$[111]_C$, which is common to all states is suppressed in Table 1.
The subindex $X$ refers to the spatial, $S$ to the spin,
$F$ to the flavor and $FS$ to the intermediate
$SU(6)_{FS}\supset SU(3)_F\times SU(2)_S$ symmetry. The latter
is required for the unique characterization of the state. The
Pauli principle requires antisymmetry with respect to permutation
of all quark coordinates - i.e. the an overall permutation
symmetry $[111]_{XCSF}$. Note that as the color-($[111]_C$),
spin- ($[21]_S$)  and
color-spin-
symmetries ($[21]_{CS}$) of the nucleon and $\Lambda$
states listed in Table 1 are identical
the color-magnetic interaction (2) cannot split them in
different directions.\\

The effect of the chiral field interaction (1) on the $J=\frac{1}{2}$
in Table 1 can be calculated by algebraic methods if the difference
in the radial structure of the $[3]_X$ and $[21]_X$ baryon states
is neglected. That would in fact be an exact result for
the totally symmetric spatial state, for which the spatial
and flavor-spin matrix elements factorize completely,
and should be a sufficiently adequate approximation
for the present purposes. As the spin-flavor part of (1) is
an invariant of the $SU(6)$ group in the reduction
$SU(6)\supset SU(3)\times SU(2)$ we have, using a result originally
developed for the color-magnetic interaction [5],

$$<[f]^{SU(6)}][f]^{SU(3)}[f]^{SU(2)}]\vert \sum _{i<j}^N
\vec\lambda_i\cdot\vec \lambda_j \vec\sigma_i
\cdot\vec\sigma_j\vert
[f]^{SU(6)}[f]^{SU(3)}[f]^{SU(2)}]>$$
$$=4C_2^{(6)}-2C_2^{(3)}
-\frac{4}{3}C_2^{(2)}-8N,\eqno(3)$$

where $N$ is the number of quarks and

$$C_2^{(n)}=<[f]^{SU(n)}|C_2^{(n)}|[f]^{SU(n)}>=$$
$${1\over 2}[f'_1(f'_1+n-1)+f'_2(f'_2+n-3)+...$$
$$+f'_{n-1}(f'_{n-1}-n+3)]-
\frac{1}{2n}(\sum_{i=1}^{n-1}f'_i)^2.\eqno(4)$$
Here $n$ is rank of the corresponding group,
$f'_i\equiv f_i-f_n$ and $f_i$ denotes the length of the $i-th$
row of the corresponding Young pattern. With this result we find that
the matrix elements (3) for the states $|[21]_{SF}[21]_F[21]_S>$,
$|[21]_{SF}[111]_F[21]_S>$, $|[3]_{SF}[21]_F[21]_S>$
and $|[21]_{SF}[21]_F[3]_S>$
are
$2$, $8$, $14$ and $-2$ respectively. These are the only ones
required for the baryon states in Table 1.\\

The effective coupling constant $C_\chi$ in the chiral interaction
(1), which represents an averaged radial matrix element, may be
determined by the mass difference
of any two baryon states that have the same radial
structure in the $SU(3)$ harmonic oscillator model (see below),
the same spin-
flavor symmetry but different spin- or (and) flavor symmetries.
Such are eg. the nucleon and the $\Delta_{33}$,
the $\Lambda(1405)$ and
$\Sigma(1750)$ resonances, the $\Sigma$ and the $\Sigma(1385)$
and the $\Xi$ and the $\Xi(1530)$. The values for $C_\chi$
obtained by these mass differences are 29.3, 34.5, 19.0 and 21.2
MeV
respectively. The spread in these values is a
reflection of the neglect of the orbital structure of the
states and the explicit flavor dependence of averaged matrix
element $C_\chi$ in (1).
We shall choose for $C\chi$ the average of these values:

$$C_\chi\sim26\, {\rm MeV}.\eqno(5)$$

In order to estimate the mass differences between the baryon states
with different orbital excitation we adopt the $SU(3)$ oscillator
version of the constituent quark model, in which the confining
interaction is harmonic. In this model the oscillator
parameter $\hbar\omega$
may be extracted from the mass differences between the first excited
$\frac{1}{2}^+$ states and the ground states of the baryons as long
as the radial structure of the interaction (1)
is neglected. This yields $\hbar\omega
\sim 250$ MeV.
In the $N$ and $\Sigma$ sectors the mass
difference between the excited ${1\over 2}^+$ and ${1\over 2}^-$
states in Table 1 will then be

$$N:\quad m({1\over 2}^+)-m({1\over 2}^-)=250\, {\rm
MeV}-C_\chi(14-2)$$
$$=-62\, {\rm MeV},\eqno(6a)$$
$$\Sigma:\quad m({1\over 2}^+)-m({1\over 2}^-)=250\, {\rm
MeV}-C_\chi(14+2)$$
$$=-166\, {\rm MeV},\eqno(6b)$$

whereas it for the $\Lambda$ system should be

$$\Lambda:\quad m({1\over 2}^+)-m({1\over 2}^-)=250\, {\rm
MeV}-C_\chi(14-8)$$
$$=94\, {\rm MeV}. \eqno(7)$$
For a lowest order estimate
these numbers agree well enough with the empirical values -95 MeV,
-90 MeV and
+195 MeV respectively. The agreement with the last value improves
if it is replaced by the difference of 137 MeV between the
$\Lambda(1600)$ and the average mass of the $\Lambda(1520)$
and $\Lambda(1405)$ as would be proper because of the
neglect of the spin-orbit interactions between the quarks.
That the $SU(2)_I\times SU(2)_S$
version of the chiral field interaction (1)
may be of importance for the explanation of the
splitting between the positive and negative parity nucleon
resonances has in fact been noted earlier [6].
In fact this represents
an explanation of
the whole low-lying part of the baryon spectrum where all
resonance masses are within 10\% of their empirical
values where known. It is readily seen
that if the parameter $C_\chi$ in (1)
is allowed to be different in the nucleon and strange hyperon
sectors, as one would expect it to be by the considerable
empirical mass splitting within the octet of Goldstone bosons,
the fit to the empirical spectra can be much improved
above the present qualitative level.
The present
description
of the baryon spectrum should of
course also be refined by a proper treatment of the anharmonicity
of the confining potential, which is linear in r rather than
quadratic, as well as by consideration of
the spatial behavior of the chiral field
interaction and the
mass
difference between the constituent u,d and s quarks [3,4].\\

A short digression on the sign of the chiral interaction (1) is in
order. This corresponds to that of the usual pion exchange potential
at short distances, where the interaction is attractive
in completely symmetric spin-isospin states
and repulsive in antisymmetric states (the relevant
term is either represented
as a $\delta$-function or as a regularizing term, which
dominates at short ranges $\leq 1$ fm).
This should be dominant for
the baryon states, which are confined within ranges  of $\sim  0.8 $fm.
The argument generalizes directly to $SU(3)_F$. In that case one has
the matrix elements

$$<[f_{ij}]_F\times [f_{ij}]_S|\vec \lambda^F_i \cdot \vec \lambda_j^F
\vec\sigma_i \cdot \vec \sigma_j|[f_{ij}]_F \times [f_{ij}]_S>$$
$$=\left\{\begin{array}{rr} {4\over 3} & [2]_F,[2]_S:[2]_{FS} \\ 8 &
[11]_F,[11]_S:[2]_{FS} \\ -4 & [2]_F,[11]_S:[11]_{FS}\\ -{8\over
3} & [11]_F,[2]_S:[11]_{FS}\end{array}\right..\eqno(8)$$

Symmetrical $FS$ pair states thus experience an attractive interaction
at short range, whereas antisymmetrical ones experience repulsion.
This explains why the $[3]_{SF}$ state in the $N(1440)$ and
$\Sigma(1660)$ positive parity resonances experiences a much larger
attractive interaction than the mixed symmetry state $[21]_{SF}$ in the
$N(1535)$ and $\Sigma(1750)$ resonances. Consequently the masses of the
$J^P={1\over 2}^+$ states $N(1440)$ and $\Sigma(1660)$ are shifted
down relative to the two other ones, which explains the reversal of
the otherwise expected "normal ordering". \\

The situation is different in the case of the $\Lambda(1405)$ and
$\Lambda(1600)$, as the flavor state of the $\Lambda(1405)$ is
totally antisymmetric. Because of this, even with the mixed $[21]_{SF}$
state, the $\Lambda(1405)$ experiences a gain in
attractive energy, which is
comparable to that of the $\Lambda(1600)$ and thus the ordering
suggested by the confining oscillator interaction is not reversed.\\

The chiral interaction (1) also, of course, predicts the ordering of
the of the strangeness -2 and -3 hyperon states.
If the mass of the light and strange constituent
quarks were equal the ordering of the $\Sigma$ and  $\Xi$
hyperons should coincide with that in the combined nucleon
and $\Delta$ spectrum.
As a consequence the $\Xi(1690)$ would be a ${1\over 2}^+$ state.
The ordering of the spectrum of
the $\Omega^-$ should be the same as that of the $\Delta$
and hence the $\Omega^-(2250)$ should be a ${3\over
2}^+$ state. \\

These results suggest that the baryon spectrum can be understood in
the following way. At low energies the chiral symmetry of the
underlying QCD is realized in the hidden mode, and the spectrum
consists of states without nearby parity partners. In the nucleon case
this part of the spectrum is formed by the nucleon and the
$\Delta(1232)$
and in the
$S=-1, I=0$ sector it is formed by the $\Lambda$, and the
negative parity states
$\Lambda(1405)$ and $\Lambda(1520)$. In this low energy region
the gross
structure of the spectrum is caused by the confining
interaction, and (most of) the fine structure by the
interaction (1)
that is mediated by the octet of pseudoscalar Goldstone bosons, which are
associated with the hidden mode of chiral symmetry. The high energy
part of the baryon spectrum on the other hand, which is formed of near
degenerate parity doublets (or more generally multiplets), reveals the
explicit Wigner-Weyl mode of chiral symmetry, which is due to the
indistinguishability between left- and right-handed massless quarks
in QCD. The remaining small splitting of the degeneracy between the
parity partners is then due to the small mass
of the current quarks and the gradually vanishing hidden mode of
chiral symmetry.
This is so as the dynamical masses of the constituent quarks
arise as a consequence of the spontaneous breaking of chiral
symmetry [7,8].
Similarly the shell gap $\hbar\omega$ that arises from the
confining force in
the oscillator model
becomes obscured by the chiral interaction (1),
which mixes the states in adjacent oscillator shells.\\

The baryon spectrum suggests that the
phase transition between
the Nambu-Goldstone and Wigner-Weyl mode of chiral symmetry is
gradual, as the mass difference between the nearest neighbours with
opposite parity falls to zero only gradually with increasing
resonance energy. The clearest signal for this is that while
the
splitting within the $\Lambda(1600)-\Lambda(1670)$ parity
doublet is still 70 MeV, the splittings within the $J^P=
\frac{1}{2}^\pm$ and $J^P=\frac{5}{2}^\pm$ $\Lambda$-resonance
parity doublets
around 1800 MeV are only 10 MeV.
This is an indication of the amorfic (disordered) structure of
the QCD vacuum and its quark condensate. The implication
would then be that
there is a
gradual chiral restoration phase transition.
The gradual nature of the phase transition also appears in
the instanton [9] liquid model of the QCD vacuum [8,10].
If
the onset
of the parity doubling in the resonance spectrum is taken to be
at about
500 MeV above the ground state, as suggested by the mass difference
between the $\Lambda$ and the lowest parity doublet formed by the
$\Lambda(1600)$ and the $\Lambda(1670)$,
the approximate transition energy would
then by of the order 500 MeV.
The absence of
structure in the baryon spectrum above 2 GeV excitation energy
shows that both
the Nambu-Goldstone mode and confinement have totally
disappeared in that
energy range.\\

The present results indicate that the role of the color-magnetic
interaction (2), which is arises from gluon exchange and
which should be important in the Wigner-Weyl mode,
for the ordering of the baryon spectrum is small. If
it is included in the model as a phenomenological term the value of
the effective coupling strength $\alpha_S$ should be much smaller than
the values $\sim 1$ that have been typically employed [3,4]. The present
suggestion for the decoding of the baryon spectrum is of course very
different from that in refined versions of the chiral bag model, that
attempt to the explain the
fine structure in terms of the color-magnetic model,
and which can predict the low mass of the $\Lambda(1405)$ only with
the extreme conclusion that it has almost no flavor singlet
component [11]. Closer in spirit, are the extended
Nambu-Jona-Lasinio [12] and chiral soliton models [13,14] in which the
low mass of the $\Lambda(1405)$ resonance also
emerges naturally. In these the Pauli principle
at the quark level is however missing
and the interaction between the constituents is
mediated by (implicit)
heavy meson exchanges rather than by pseudoscalar bosons.\\

It should finally be pointed out that the chiral field
interaction (1) is not expected to be important in the heavy
flavor charm and bottom sectors, because of the large breaking
of chiral symmetry caused by the large masses of the charm
and bottom quarks. Further application of the the pseudoscalar
octet mediated interaction should take into account
the flavor symmetry breaking caused by the mass splitting
within the octet as well as
the spatial dependence of the interaction,
which - at least in the case of
two-
baryon systems - should be important [15,16].\\

\vspace{2cm}

{\bf Acknowledgements}
\vspace{0.5cm}

This research has been supported by Academy of Finland Grant \#7635.

\newpage

{\bf References}
\vspace{0.5cm}

\begin{enumerate}
\item S. Weinberg, Phys.Rev. {\bf D11}. 3583 (1975)
\item A. DeRujula, H. Georgi and S.L. Glashow, Phys.Rev.
{\bf D12} (1975) 147
\item N. Isgur and G. Karl, Phys. Rev. {\bf D18}, 4187 (1978)
\item N. Isgur and G. Karl, Phys. Rev. {\bf D19}, 2653 (1979)
\item R.-L. Jaffe, Phys. Rev. Lett. {\bf 38}, 195 (1977)
\item I.T. Obukhovsky and A.M. Kusainov, Phys. Lett. {\bf B238}, 142
(1990)
\item A. Manohar and H. Georgi, Nucl.Phys. {\bf B234}, 189 (1984)
\item D.I. Diakonov and V.Yu. Petrov, Nucl.Phys. {\bf B272}. 457 (1986)
\item A.A. Belavin, A.M. Polyakov, A.S. Schwartz and
Yu.S. Tyapkin,
Phys.Lett. {\bf 59B}, 85 (1975)
\item E.V. Shuryak, Phys.Rep. {\bf C115}, 152 (1984) and references
therein
\item Y. Umino and F. Myhrer, Phys. Rev. {\bf D39}, 3391 (1989)
\item H. Weigel, R. Alkhofer and H. Reinhardt,
 Phys. Rev. {\bf D49}, 5958 (1994)
\item C.G. Callan, K. Hornbostel and I. Klebanov, Phys. Lett. {\bf
B202}, 269 (1988)
\item U. Blom, K. Dannbom and D.O. Riska, Nucl. Phys. {\bf A493}, 384
(1989)
\item L.Ya. Glozman and E.I. Kuchina, Phys. Rev. {\bf C49}, 1149 (1994)
\item Z. Zhang, A. Faessler, U. Straub and L.Ya. Glozman, Nucl. Phys.
{\bf A578}, 573 (1994)
\end{enumerate}

\newpage
\centerline{\bf Table 1}
\vspace{0.5cm}

The structure of the lowest confirmed $J={1\over 2}$ baryon
states

\begin{center}
\begin{tabular}{|lll|} \hline
$N$ & $N(1440)$ & $N(1535)$\\ \hline
${1\over 2}^+$ & ${1\over 2}^+$ & ${1\over 2}^-$\\
$[3]_X$ & $[3]_X$ & $[21]_X$\\
$[3]_{SF}$ & $[3]_{SF}$ & $[21]_{SF}$\\
$[21]_F$ & $[21]_F$ & $[21]_F$\\
$[21]_S$ & $[21]_S$ & $[21]_S$\\ \hline
\end{tabular}
\end{center}
\vspace{0.3cm}

\begin{center}
\begin{tabular}{|lll|} \hline
$\Lambda$ & $\Lambda(1405)$ & $\Lambda(1600)$ \\ \hline
${1\over 2}^+$ & ${1\over 2}^-$ & ${1\over 2}^+$\\
$[3]_X$ & $[21]_X$ & $[3]_X$\\
$[3]_{SF}$ & $[21]_{SF}$ & $[3]_{SF}$\\
$[21]_F$ & $[111]_F$ & $[21]_F$\\
$[21]_S$ & $[21]_S$ & $[21]_S$\\ \hline
\end{tabular}
\end{center}
\vspace{0.3cm}

\begin{center}
\begin{tabular}{|lll|} \hline
$\Sigma$ & $\Sigma(1660)$ & $\Sigma(1750)$\\ \hline
${1\over 2}^+$ & ${1\over 2}^+$ & ${1\over 2}^-$\\
$[3]_X$ & $[3]_X$ & $[21]_X$\\
$[3]_{SF}$ & $[3]_{SF}$ & $[21]_{SF}$\\
$[21]_F$ & $[21]_F$ & $[21]_F$\\
$[21]_S$ & $[21]_S$ & $[3]_S$\\ \hline
\end{tabular}
\end{center}

\end{document}